\documentclass[12pt]{article}

\usepackage[dvips]{graphicx}
\usepackage{latexsym}

\usepackage{amssymb,amsfonts,amsmath}
\usepackage{graphicx} 
\usepackage{indentfirst}

 \usepackage{bbm}

\parskip=\medskipamount

\newcommand {\cC}{{\cal C}}
\newcommand {\cD}{{\cal D}}

\newcommand {\cF}{{\cal F}}

\newcommand {\cL}{{\cal L}}
\newcommand {\cM}{{\cal M}}
\newcommand {\cN}{{\cal N}}

\def\a{\alpha}
\def \bi{\bibitem}

\def\b{\beta}

\def\d{\delta}
\def\e{\epsilon}

\def\g{\gamma}
\def\G{\Gamma}

\def\k{\kappa}
\def\l{\lambda}
\def\m{\mu}
\def\n{\nu}

\def\q{\theta}

\def\s{\sigma}

\def\x{\xi}

\def\D{\Delta}
\def\F{\Phi}

\def\L{\Lambda}
\def\O{\Omega}

\def\S{\Sigma}
\def\U{\Upsilon}
\def\X{\Xi}

\def\rd{{\rm d}}
\def\ri{{\rm i}}

\newcommand{\ve}{\varepsilon}                            

\newcommand{\pa}{\partial}                           
\newcommand{\hf}{\frac12}

\newcommand{\sect}[1]{\setcounter{equation}{0}\section{#1}}

\newcommand{\be}{\begin{equation}}
\newcommand{\ee}{\end{equation}}
\newcommand{\bea}{\begin{eqnarray}}
\newcommand{\eea}{\end{eqnarray}}
\newcommand{\non}{\nonumber}
\newcommand{\1}{\underline{1}}
\newcommand{\2}{\underline{2}}

\newcommand{\bm}[1]{\mbox{\boldmath$#1$}}

\def\double #1{#1{\hbox{\kern-2pt $#1$}}}

\newcommand{\hm}{{\hat{m}}}

\newcommand{\ha}{{\hat{a}}}
\newcommand{\hb}{{\hat{b}}}
\newcommand{\hc}{{\hat{c}}}
\newcommand{\hd}{{\hat{d}}}
\newcommand{\he}{{\hat{e}}}

\newcommand{\hal}{{\hat{\a}}}
\newcommand{\hbe}{{\hat{\b}}}
\newcommand{\hga}{{\hat{\g}}}
\newcommand{\hde}{{\hat{\d}}}

\begin{document}

\begin{titlepage}

\begin{flushright}
February, 2008\\
\end{flushright}
\vspace{5mm}

\begin{center}
{\Large \bf  Super-Weyl invariance in 5D supergravity }\\ 
\end{center}

\begin{center}

{\large  
Sergei M. Kuzenko\footnote{{kuzenko@cyllene.uwa.edu.au}}
and 
Gabriele Tartaglino-Mazzucchelli\footnote{gtm@cyllene.uwa.edu.au}
} \\
\vspace{5mm}

\footnotesize{
{\it School of Physics M013, The University of Western Australia\\
35 Stirling Highway, Crawley W.A. 6009, Australia}}  
~\\

\vspace{2mm}

\end{center}
\vspace{5mm}

\begin{abstract}
\baselineskip=14pt
We propose a superspace formulation for the  Weyl multiplet of $\cN=1$ conformal
supergravity in five dimensions. 
The corresponding superspace constraints are invariant 
under super-Weyl transformations generated by a real scalar parameter. 
The minimal supergravity multiplet, which was
introduced by Howe in 1981, emerges if one  couples the Weyl multiplet 
to an Abelian vector multiplet and then breaks the super-Weyl invariance 
by imposing the gauge condition $W=1$, with $W$ the  field strength
of the vector multiplet.
The geometry of superspace is shown to allow the existence of a large family of off-shell 
supermultiplets that  possess uniquely determined super-Weyl transformation laws
and can be used to describe supersymmetric matter.
Many of these supermultiplets have not appeared within the superconformal 
tensor calculus. We formulate a manifestly locally supersymmetric and super-Weyl invariant  
action principle. In the super-Weyl gauge $W=1$, this  action reduces to that constructed in 
 arXiv:0712.3102. We also  present a superspace formulation for the dilaton Weyl 
 multiplet.
\end{abstract}
\vspace{1cm}

\vfill
\end{titlepage}

\newpage
\renewcommand{\thefootnote}{\arabic{footnote}}
\setcounter{footnote}{0}

\tableofcontents{}
\vspace{1cm}
\bigskip\hrule


\sect{Introduction}

Recently, we have constructed a  superspace formulation 
for general  $\cN=1$ (often called $\cN=2$) supergravity-matter systems
\cite{KT-Msugra5D,KT-Msugra5D2} in five space-time dimensions.
In the approach of \cite{KT-Msugra5D,KT-Msugra5D2},
the geometry of curved superspace is described by the minimal supergravity multiplet 
introduced by Howe in 1981 \cite{Howe5Dsugra}
(see also \cite{HL}).
On the other hand, 
if one describes 5D $\cN=1$ matter-coupled supergravity\footnote{Matter couplings
in  5D $\cN=1$ supergravity have also been studied  
within the  on-shell component approaches  \cite{GST,GZ,CD}.} 
using the component superconformal tensor calculus
\cite{Ohashi,Bergshoeff}, 
the natural starting point is the  Weyl multiplet. 
In the latter setting, the minimal multiplet\footnote{The minimal supergravity multiplet
 was re-discoverd in \cite{Zucker} 
where the component implications of \cite{Howe5Dsugra} were elaborated.}
 \cite{Howe5Dsugra} occurs by coupling the Weyl multiplet 
to an Abelian vector multiplet, and then 
breaking the Weyl invariance and some other local symmetries. 
To the best of our knowledge, the Weyl multiplet has never been realized 
in superspace.\footnote{Applying the harmonic superspace approach 
\cite{GIKOS,GIOS} to 5D $\cN=1$ supergravity, it is not difficult to construct 
a prepotential realization for the Weyl multiplet. It is also not difficult 
to derive the supercurrents \cite{HL,Bergshoeff}
 (and the multiplet of anomalies) by varying the matter action with respect 
to the supergravity prepotentials, similarly to the 4D $\cN=2$ case \cite{KT}.  
It is non-trivial, however, to relate the prepotential realization 
to an underlying  covariant geometric formulation for supergravity-matter system.  
The latter formulation is elaborated in this paper.}
The present paper is aimed at filling this gap. 

Quaternion-K\"ahler spaces are known to be the target spaces
for locally supersymmetric nonlinear sigma-models with eight supercharges
\cite{BW}.  As is known, there exists a one-to-one
correspondence between $4n$-dimensional quaternion-K\"ahler spaces
and $4(n+1)$-dimensional hyperk\"ahler manifolds possessing a homothetic 
Killing vector (implying the fact that the isometry group 
includes a subgroup SU(2) that rotates  the three complex structures)
\cite{Swann,Galicki}.  In the physics literature, such hyperk\"ahler spaces
are known as ``hyperk\"ahler cones'' \cite{deWRV}. They emerge as the target spaces
for rigid  superconformal sigma-models with eight supercharges in diverse dimensions
(see \cite{deWRV} and references therein). 
The analysis in \cite{deWRV} shows that in order 
to generate quaternion-K\"ahler metrics from hyperk\"ahler cones, 
one essentially needs two prerequisites: (i) a superspace formulation 
for general rigid superconformal sigma-models with eight supercharges;
(ii) a superspace extension of the superconformal tensor calculus. 
General rigid superconformal multiplets and their sigma-models couplings
in projective superspace \cite{KLR,LR,G-RLRvUW} 
have been given in \cite{K,K2}
in five and four space-time dimensions.
The present paper provides  the desired superspace extension 
of the superconformal tensor calculus in the case of five dimensions. 
The case of 4D $\cN=2$ supergravity will be considered elsewhere \cite{KLRT}.

This paper is organized as follows. In section 2 we derive a superspace 
formulation for the standard Weyl multiplet in which the super-Weyl 
transformations are generated by an unconstrained real parameter.
Section 3 is devoted to an alternative formulation in which the super-Weyl 
transformations are generated by a constrained real parameter.
We also provide a superspace realization for the dilaton Weyl multiplet 
\cite{Ohashi,Bergshoeff} that corresponds to the Nishino-Rajpoot version 
\cite{NR}
of 5D $\cN=1$ Poincar\'e supergravity.
In section 4 we introduce a large family of off-shell 
supermultiplets that  possess uniquely determined super-Weyl transformation laws
and can be used to describe supersymmetric matter. Finally, in section 5 
we present a  manifestly locally supersymmetric and super-Weyl invariant  
action principle. 


\sect{The Weyl multiplet in superspace}

Let $z^{\hat{M}}=(x^{\hm},\q^{\hat{\mu}}_i)$
be local bosonic ($x$) and fermionic ($\q$) 
coordinates parametrizing  a curved five-dimensional $\cN=1$  superspace
$\cM^{5|8}$,
where $\hm=0,1,\cdots,4$, $\hat{\mu}=1,\cdots,4$, and  $i=\1,\2$.
The Grassmann variables $\q^{\hat{\mu}}_i$
are assumed to obey the standard pseudo-Majorana reality condition
$(\q^{\hat{\mu}}_i)^* = \q_{\hat{\mu}}^i =\ve_{\hat{\m} \hat{\n}}\,  \ve^{ij} \, \q^{\hat{\nu}}_j  $
(see the appendix of \cite{KT-Msugra5D2} for our 5D  notation and conventions).
The tangent-space group
is chosen to be  ${\rm SO}(4,1)\times {\rm SU}(2)$,
and the superspace  covariant derivatives 
$\cD_{\hat{A}} =(\cD_{\hat{a}}, \cD_{\hat{\a}}^i)$
have the form 
\bea
\cD_{\hat{A}}&=&
E_{\hat{A}} + \O_{\hat{A}} + \F_{\hat{A}}
~.
\label{CovDev}
\eea
Here $E_{\hat{A}}= E_{\hat{A}}{}^{\hat{M}}(z) \,\pa_{\hat{M}}$ is the supervielbein, 
with $\pa_{\hat{M}}= \pa/ \pa z^{\hat{M}}$,
\bea
\O_{\hat{A} }= \hf \,\O_{\hat{A}}{}^{\hb\hc}\,M_{\hb\hc}
= \O_{\hat{A}}{}^{\hbe\hga}\,M_{\hbe\hga}~,\qquad 
M_{\ha\hb}=-M_{\hb\ha}~, \quad M_{\hal\hbe}=M_{\hbe\hal}
\eea
is the Lorentz connection, 
\bea
\F_{\hat{A}} = \F^{~\,kl}_{\hat{A}}\,J_{kl}~, \qquad
J_{kl}=J_{lk}
\eea
is the SU(2)-connection. 
The Lorentz generators with vector indices ($M_{\ha\hb}$) and spinor indices
($M_{\hal\hbe}$) are related to each other by the rule:
$M_{\ha\hb}=(\S_{\ha\hb})^{\hal\hbe}M_{\hal\hbe}$ 
(for more details, see the appendix of \cite{KT-Msugra5D2}).
The generators of ${\rm SO}(4,1)\times {\rm SU}(2)$
act on the covariant derivatives as follows:\footnote{The operation of
(anti)symmetrization of $n$ indices 
is defined to involve a factor $(n!)^{-1}$.}
\bea
{[}J^{kl},\cD_{\hal}^i{]}
= \ve^{i(k} \cD^{l)}_{\hat \a}~,~~~
{[}M_{\hal\hbe},\cD_{\hga}^k{]}
=\ve_{\hga(\hal}\cD^k_{\hbe)}~,~~~
{[}M_{\ha\hb},\cD_{\hc}{]}
=2\eta_{\hc[\ha}\cD_{\hb]}~,
\label{generators}
\eea
where $J^{kl} =\ve^{ki}\ve^{lj} J_{ij}$.

The supergravity gauge group is generated by local transformations
of the form 
\be
\d_K \cD_{\hat{A}} =[ K, \cD_{\hat{A}} ]~,
\qquad K = K^{\hat{C}}(z) \cD_{\hat{C}} +\hf K^{\hat c \hat d}(z) M_{\hat c \hat d}
+K^{kl}(z)J_{kl}
~,
\label{tau}
\ee
with all the gauge parameters 
obeying natural reality conditions, and otherwise  arbitrary. 
Given a tensor superfield $U(z)$, with its indices suppressed, 
it transforms as follows:
\bea
\d_K U = {K  }\, U~.
\eea

The covariant derivatives obey (anti)commutation relations of the general form 
\bea
{[}\cD_{\hat{A}},\cD_{\hat{B}}\}&=&T_{\hat{A}\hat{B}}{}^{\hat{C}}\cD_{\hat{C}}
+\hf R_{\hat{A}\hat{B}}{}^{\hat{c}\hat{d}}M_{\hat{c}\hat{d}}
+R_{\hat{A}\hat{B}}{}^{kl}J_{kl}
~,
\label{algebra}
\eea
where $T_{\hat{A}\hat{B}}{}^{\hat{C}}$ is the torsion, 
$R_{\hat{A}\hat{B}}{}^{kl}$ and $R_{\hat{A}\hat{B}}{}^{\hat{c}\hat{d}}$  
the SU(2)- and SO(4,1)-curvature tensors, respectively. 

\subsection{Constrained superspace geometry}
We choose the torsion to obey the constraints:
\begin{subequations}
\bea
T_{\hal}^i{}_{\hbe}^j{}^{\hc}~=~-2\ri\ve^{ij}(\Gamma^{\hc})_{\hal\hbe}  \qquad
\qquad && \qquad  
\mbox{(dim 0)}
\label{constr-0} \\
T_{\hal}^i{}_{\hbe}^j{}^{\hga}_k~=~
T_{\hal}^i{}_{\hb}{}^{\hc}~
=~0  \qquad  \qquad && \qquad 
\mbox{(dim $\frac{1}{2}$)}
\label{constr-1/2} \\
T_{\ha\hb}{}^{\hc}~=~T_{\ha\hbe (j}{}^{\hbe}{}_{k)}~=~0  \qquad
\qquad && \qquad  
\mbox{(dim 1)}~.
\label{constr-1}
\eea
\end{subequations}
The set of constraints (\ref{constr-0} -- \ref{constr-1})  is obtained from that 
defining the minimal supergravity multiplet \cite{Howe5Dsugra}
by removing those constraints which correspond to 
the central-charge field strength. 

With  the constraints introduced,  it can be shown that
the torsion and the curvature tensors in (\ref{algebra}) 
are expressed in terms of a small number of 
dimension-1 tensor superfields, $S^{ij}$, $X_{\ha\hb}$, 
$N_{\ha\hb}$ and $C_\ha{}^{ij}$, and their covariant derivatives,  
with the symmetry properties:
\bea
S^{ij}=S^{ji}~,\qquad X_{\ha\hb}=-X_{\hb\ha}~,\qquad N_{\ha\hb}=-N_{\hb\ha}~,
\qquad C_\ha{}^{ij}=C_\ha{}^{ji}~.
\eea
Their reality properties are
\be
\overline{S^{ij} } =S_{ij}~, \qquad 
\overline{X_{\ha\hb}} =X_{\ha\hb}~, \qquad 
\overline{N_{\ha\hb}} =N_{\ha\hb}~, \qquad
\overline{C_\ha{}^{ij} }=C_{\ha ij }~.
\ee
The covariant derivatives obey the (anti)commutation relations:
\begin{subequations}
\bea
\big\{ \cD_{\hal}^i , \cD_{\hbe}^j \big\} &=&-2 \ri \,\ve^{ij}\cD_{\hal\hbe}
-\ri \,\ve_{\hal\hbe}\ve^{ij}X^{\hc\hd}M_{\hc\hd}
+{\ri\over 4} \ve^{ij}\ve^{\ha\hb\hc\hd\he}(\G_\ha)_{\hal\hbe}N_{\hb\hc}M_{\hd\he}
\non\\
&&
-{\ri\over 2}\ve^{\ha\hb\hc\hd\he}(\S_{\ha\hb})_{\hal\hbe}C_{\hc}{}^{ij}M_{\hd\he}
+4\ri \,S^{ij}M_{\hal\hbe}
+3\ri \, \ve_{\hal\hbe}\ve^{ij}S^{kl}J_{kl}
\non\\
&&
-\ri \, \ve^{ij}C_{\hal\hbe}{}^{kl}J_{kl}
-4\ri\Big(X_{\hal\hbe}+N_{\hal\hbe}\Big)J^{ij}
~,
\label{covDev2spinor-} \\
{[}\cD_\ha,\cD_{\hbe}^j{]}&=&
{1\over 2} \Big(
(\Gamma_{\hat{a}})_{\hbe}{}^{\hga}S^j{}_k
- X_{\ha\hb}(\Gamma^{\hat{b}})_{\hbe}{}^{\hga} \d^j_k
-{1\over 4}\,\ve_{\ha\hb\hc\hd\he}N^{\hd\he}(\Sigma^{\hb\hc})_{\hbe}{}^{\hga}
\d^j_k
+ (\S_\ha{}^{\hb})_{\hbe}{}^{\hga}C_\hb{}^j{}_k
\Big)
\cD_{\hga}^k
\non\\
&&
-{\ri\over 2}\Big((\G_\ha)_{\hbe}{}^\hga T^{\hc\hd}{}_\hga^{j}
+2(\G^{[\hc})_{\hbe}{}^\hga T_{\ha}{}^{\hd]}{}_\hga^{j}
\Big)M_{\hc\hd}
\non\\
&&
+\Big(3\X_{\ha}{}_\hbe^{(k}\ve^{l)j}
-{1\over 3}\cC_{\ha}{}_{\hbe}^{(k}\ve^{l)j}
-{5\over 4}(\G_\ha)_{\hbe}{}^\hga\cF_\hga^{(k}\ve^{l)j}
+{1\over 4}(\G_\ha)_{\hbe}{}^\hga\cN_\hga^{(k}\ve^{l)j}
\non\\
&&~~~
+{1\over 8}(\G_\ha)_{\hbe}{}^\hga\cC_{\hga}{}^{jkl}
-{11\over 24}(\G_\ha)_{\hbe}{}^\hga\cC_{\hga}^{(k}\ve^{l)j}
\Big)
J_{kl}
~.
\label{covDev2spinor-2}
\eea
\end{subequations}
The dimension-1 components of the torsion, 
 $S^{ij}$, $X_{\ha\hb}$, 
$N_{\ha\hb}$ and $C_\ha{}^{ij}$, 
enjoy some additional differential constraints 
which follow from the Bianchi identities.
To formulate them, it is useful to introduce
the irreducible components of $\cD_\hga^kX_{\ha\hb}$ and
$\cD_\hga^kC_{\ha}{}^{ij}$
defined as follows:
\begin{subequations}
\bea
\cD_\hga^kX_{\ha\hb}&=&
W_{\ha\hb\hga}{}^k
+2(\G_{{[}\ha})_{\hga}{}^{\hde}\X_{\hb{]}\hde}
+(\S_{\ha\hb})_\hga{}^\hde \cF_{\hde}{}^k~, 
\non
\\
&&(\G^\ha)_\hal{}^\hbe\X_{\ha\hbe}{}^i=(\G^\ha)_\hal{}^\hbe W_{\ha\hb\hbe}{}^i=0
~,
\label{X-irreducible}\\
\cD_\hga^kC_{\ha}{}^{ij}&=&
\cC_\ha{}_\hga{}^{ijk}
-{2\over 3}\cC_{\ha}{}_{\hga}^{(i}\ve^{j)k}
-{1\over 2}(\G_\ha)_\hga{}^{\hde}\cC_\hde{}^{ijk}
+{1\over 3}(\G_\ha)_\hga{}^{\hde}\cC_\hde^{(i}\ve^{j)k}~,
\non
\\
&&
\cC_\ha{}_\hga{}^{ijk}=\cC_\ha{}_\hga{}^{(ijk)}~,~~
\cC_\hde{}^{ijk}=\cC_\hde{}^{(ijk)}~,~~~
(\G^\ha)_\hal{}^\hbe\cC_\ha{}_\hbe{}^{ijk}=0~.
\eea
\end{subequations}
The dimension-3/2 Bianchi identities are: 
\begin{subequations}
\bea
\cD_\hga^kN_{\ha\hb}&=&
-W_{\ha\hb\hga}{}^k
+4(\G_{{[}\ha})_{\hga}{}^{\hde}\X_{\hb{]}\hde}
+(\S_{\ha\hb})_\hga{}^\hde \cN_{\hde}{}^k~,
\label{D-N} 
\\
\cC_\ha{}_\hga{}^{ijk}&=&0~,
\label{3/2D-C}
\\
\cD_\hga^kS^{ij}&=&
-{1\over 4}\cC_{\hga}{}^{ijk}
+{5\over 12}\cC_\hga^{(i}\ve^{j)k}
+{1\over 2}\Big(3\cF_\hga^{(i}+\cN_\hga^{(i}\Big)\ve^{j)k}
~.
\label{3/2DS}
\eea
\end{subequations}
The dimension-3/2 torsion is 
\bea
T_{\ha\hb}{}_\hga^k&=&
{\ri\over 2}\cD_\hga^k X_{\ha\hb}
-{\ri\over 6}(\G_{[\ha})_\hga{}^\hde\cC_{\hb]}{}_\hde^{k}
+{\ri\over 4}(\S_{\ha\hb})_\hga{}^\hde\cC_\hde^{k}~.
\label{dim-3/2-torsion}
\eea
The irreducible components of $\cD_\hga^kN_{\ha\hb}$ are defined similarly to 
(\ref{X-irreducible}). 
In accordance with eq. (\ref{D-N}), only one of them, $ \cN_{\hde}{}^k$, 
is a new superfield, while the other two components occur in (\ref{X-irreducible}). 
It is worth pointing out  that eq. (\ref{3/2DS}) implies
\bea
\cD_\hga^{(i}S^{jk)}=-{1\over 4}\cC_\hga{}^{ijk}~.
\label{3/2Dev+S++}
\eea
The latter result  will be important for our consideration below.


\subsection{Super-Weyl transformations}

A short calculation shows that the constraints (\ref{constr-0} -- \ref{constr-1})  
are invariant under super-Weyl transformations of the form:
\begin{subequations}
\bea
\d_\s \cD_\hal^i&=&\s\cD_\hal^i+4(\cD^{\hga i}\s)M_{\hga\hal}-6(\cD_{\hal k}\s)J^{ki}~,
\label{sW1} \\
\d_\s \cD_\ha&=&
2\s\cD_\ha
+\ri(\G_\ha)^{\hga\hde}(\cD_{\hga}^{k}\s)\cD_{\hde k}
-2(\cD^\hb\s)M_{\ha\hb}
+{\ri\over 4}(\G_\ha)^{\hga\hde}(\cD_\hga^{(k}\cD_{\hde}^{l)}\s)J_{kl}
\label{sW2}
~,
\eea
\end{subequations}
where the parameter $\s(z)$ is a real unconstrained superfield. 
The components of the torsion can be seen to transform as follows:
\begin{subequations}
\bea
\d_\s S^{ij}&=&2\s S^{ij}
+{\ri\over 2}\,\cD^{\hal (i}\cD_{\hal}^{ j)}\s~,
\label{s-Weyl-Sij}\\
\d_\s C_{\ha}{}^{ij}&=&2\s C_{\ha}{}^{ij}
+{\ri}\,( \G_\ha)^{\hga\hde} \cD_{\hga}^{(i}\cD_{\hde}^{j)}\s~,
\label{C-var}\\
\d_\s X_{\ha\hb}&=&2\s X_{\ha\hb}
-{\ri\over 2}\, (\S_{\ha\hb})^{\hal\hbe}\cD_\hal^k\cD_{\hbe k}\s~,
\label{s-Weyl-X}\\
\d_\s N_{\ha\hb}&=&2\s N_{\ha\hb}
-\ri \,(\S_{\ha\hb})^{\hal\hbe} \cD_{\hal}^{k}\cD_{\hbe k}\s~.
\label{s-Wey-N}
\eea
\end{subequations}
It follows that 
\be
W_{\ha \hb} := X_{\ha\hb} -\hf  N_{\ha\hb}
\ee
transforms homogeneously, and hence it can be identified 
with a superspace generalization of the Weyl tensor.

Let us analyze  the 
supergravity multiplet introduced above.
First, it consists of the fields that constitute the covariant derivatives
(\ref{CovDev}) subject to the constraints (\ref{constr-0} -- \ref{constr-1}). 
Second, it possesses the gauge freedom (\ref{tau}),  (\ref{sW1}) and  (\ref{sW2}).
It proves  to describe the Weyl multiplet \cite{Ohashi,Bergshoeff}.
Indeed, one can choose a Wess-Zumino gauge and partially fix the super-Weyl 
gauge freedom in such a way that the remaining fields and 
the residual gauge transformations match those characteristic of 
the Weyl multiplet \cite{Ohashi,Bergshoeff}. 
In particular, it follows immediately from 
(\ref{s-Weyl-Sij}), (\ref{C-var}) and (\ref{s-Weyl-X}) that the 
$\q$-independent components of $S^{ij}$,  $ C_{\ha}{}^{ij}$ and 
$X_{\ha\hb}$ can be gauged away by super-Weyl transformations.
Such an approach is quite 
laborious. Fortunately, there is a more elegant way to prove the claim. 
It is sufficient to demonstrate that 
one ends up with the minimal supergravity multiplet \cite{Howe5Dsugra}
by coupling the above multiplet to an Abelian vector multiplet 
and then breaking the super-Weyl invariance.  This will be demonstrated in 
the remainder of this section.


\subsection{Coupling to a vector multiplet}

Let us couple the Weyl multiplet to an Abelian vector multiplet. 
The covariant derivatives should be modified as follows:
\bea
\cD_{\hat{A}} \quad \longrightarrow \quad 
{\bm \cD}_{\hat{A}}:=\cD_{\hat{A}}+ V_{\hat{A}}Z~,
\eea
with $V_{\hat{A}}(z) $ the gauge connection.
We will  interpret the generator $Z $ to be  a real central charge.
It is also necessary to impose covariant constraints on some components 
of the field strength of the vector multiplet 
as in the flat case \cite{HL} (see also \cite{Zupnik,KL}).

The covariant derivatives now satisfy the algebra
\bea
{[}{\bm \cD}_{\hat{A}},{\bm \cD}_{\hat{B}}\}&=&
T_{\hat{A}\hat{B}}{}^{\hat{C}}\, {\bm \cD}_{\hat{C}}
+\hf R_{\hat{A}\hat{B}}{}^{\hc\hd}M_{\hc\hd}
+R_{\hat{A}\hat{B}}{}^{kl}J_{kl}
+F_{\hat{A}\hat{B}}Z~,
\eea
where the torsion and curvature are the same as before 
and the central charge field strengths are
\begin{subequations}
\bea
&&F_\hal^i{}_\hbe^j=-2\ri\ve^{ij}\ve_{\hal\hbe}W~,~~~
F_{\ha}{}_\hbe^j=(\G_\ha)_\hbe{}^{\hga}\cD_{\hga}^jW~,
\label{FSa}\\
&&F_{\ha\hb}=
X_{\ha\hb}W
+{\ri\over 4}(\S_{\ha\hb})^{\hga\hde}\cD_{\hga}^k\cD_{\hde k}W~.
\label{FSb}
\eea
\end{subequations}
Here the field strength $W$  is real, $\bar W = W$, and obeys the Bianchi identity
\bea
\cD_{\hal}^{(i}\cD_{\hbe}^{j)}W
-{1\over 4}\ve_{\hal\hbe}\cD^{\hga(i}\cD_{\hga}^{j)} W 
= {\ri\over 2} C_{\hal\hbe}{}^{ij} W ~.~~~~~~
\label{W-BI}
\eea

The field strength $W$ possesses the following super-Weyl transformation:
\bea
\d_\s W=2\s W~.
\label{super-Weyl-W}
\eea
It is a simple calculation to demonstrate that eq. (\ref{W-BI}) is invariant 
under the super-Weyl transformations.

\subsection{The minimal multiplet}
Suppose that the field strength of the vector multiplet is 
everywhere non-vanishing, $\langle W \rangle \neq 0$,
that is
the body of $W(z) \neq0$ for any point $z \in \cM^{5|8}$.
Then,  the super-Weyl symmetry can be used to choose the gauge
\be
W=1~.
\label{W=1}
\ee
 Now, eq. (\ref{W-BI}) reduces to 
 \bea
 C_{\ha}{}^{ij}=0~,
\label{W-BI2}
\eea
while eqs. (\ref{FSa}) and (\ref{FSb}) 
turn into
\bea
&&F_\hal^i{}_\hbe^j=-2\ri\ve^{ij}\ve_{\hal\hbe}~,\qquad
F_{\ha}{}_\hbe^j= 0~, \qquad 
F_{\ha\hb}=
X_{\ha\hb}~.
\label{FS2}
\eea
As a result, one ends up with the minimal supergravity multiplet 
\cite{Howe5Dsugra}.
This is analogous to the situation in 4D $\cN=2$ supergravity
\cite{Howe,Muller}.

\sect{Variant  formulations for the Weyl multiplet}
The fact that the super-Weyl gauge freedom allows one to gauge away $C_{\ha}{}^{ij}$,
eq. (\ref{W-BI2}), is equivalent to the existence of an alternative formulation 
for the Weyl multiplet. 

\subsection{Reduced formulation}

Let us start again from the superspace formulation for the Weyl multiplet 
we have developed in the previous section.
We are in a position  to  choose the super-Weyl gauge
(\ref{W-BI2}). 
This is equivalent to the replacement of the dimension-1 constraints (\ref{constr-1})
with 
\bea
T_{\ha\hb}{}^{\hc}~=~T_{\ha\hbe (j}{}^{\hbe}{}_{k)}~=~
T_\ha{}_{(\hbe}^{(j}{}_{\hga)}^{k)}~=~
0 ~. 
\label{constr-1-mod}
\eea
Then, eq. (\ref{3/2Dev+S++})
turns into 
\bea
\cD_{\hal}^{(i}S^{jk)} =0~.
\label{S-Bianchi}
\eea

In accordance with (\ref{C-var}), the residual super-Weyl transformations 
are generated by a  parameter obeying the constraint
\bea
\cD_{\hal}^{(i}\cD_{\hbe}^{j)}\s
-{1\over 4}\ve_{\hal\hbe}\cD^{\hga(i}\cD_{\hga}^{j)} \s =0~.
\label{s-Weyl-rest}
\eea
Clearly, this (partially gauged-fixed) superspace setting still describes the Weyl multiplet. 
However, the present formulation is technically much simpler to deal with 
than the one developed in the previous section.
In what follows, we will only use the formulation 
for the Weyl multiplet which is given in the present section.
It will be referred to as the {\it reduced formulation}. 

\subsection{Coupling to a vector multiplet}
If an Abelian vector multiplet is coupled to the Weyl multiplet, 
 the corresponding field strength $W$ obeys the Bianchi identity 
\bea
\cD_{\hal}^{(i}\cD_{\hbe}^{j)}W
-{1\over 4}\ve_{\hal\hbe}\cD^{\hga(i}\cD_{\hga}^{j)} W =0
~~~~~~~
\label{W-BI-2}
\eea
which is  obtained from (\ref{W-BI}) by setting $C_{\hal\hbe}{}^{ij}=0$.
Comparing (\ref{s-Weyl-rest}) with (\ref{W-BI-2}), we see 
that $W$ and the super-Weyl parameter are constrained superfields
of the same type. 

Le us consider the composite superfield
\bea
G^{ij}:=\ri\,\cD^{\hal (i}W\cD_\hal^{j)} W+{\ri\over 2} W \cD^{ij} W-2S^{ij}W^2~,
\qquad \cD^{ij}:= \cD^{\hal (i} \cD_\hal^{j)} ~,
\label{Gij}
\eea
which is a curved superspace extension of the composite 
$O(2)$ multiplet  introduced in \cite{KL} and later in \cite{KT-M}
for the flat and Anti-de Sitter cases, respectively.
Its crucial property is 
\bea
\cD^{(i}_\hal G^{jk)}=0~.
\label{G-anal}
\eea
The super-Weyl transformation of $G^{ij}$ can be shown, 
with the use of eqs. (\ref{sW1}) and  (\ref{super-Weyl-W}),
to be 
\bea
\d_\s G^{ij}&=&6\s G^{ij}~.
\label{super-Weyl-G}
\eea

If the field strength $W$ is everywhere non-vanishing,  $\langle W \rangle \neq 0$,
the super-Weyl gauge freedom can be used to choose 
the gauge (\ref{W=1}). The   result is  again    the minimal supergravity multiplet 
\cite{Howe5Dsugra}.

\subsection{The dilaton Weyl multiplet}
In superspace, the dilaton Weyl multiplet \cite{Ohashi,Bergshoeff} 
can be  realized as the standard  Weyl
multiplet coupled to an Abelian vector multiplet  such that 
its  field strength $W$ is everywhere non-vanishing, 
$\langle W \rangle \neq 0$, and enjoys the equation
\bea
G^{ij}
=\ri\,\cD^{\hal (i}W\cD_\hal^{j)} W+{\ri\over 2} W \cD^{ij} W-2S^{ij}W^2
=0~.
\label{Chern-Simons}
\eea
The latter is equivalent to
\be
S^{ij} = \frac{\rm i}{2W^2} \Big\{  
\cD^{\hal (i}W\cD_\hal^{j)} W+\hf W \cD^{ij} W\Big\}~.
\ee
Similarly to the rigid supersymmetric case \cite{KL}, 
eq. (\ref{Chern-Simons}) originates as the equation of motion 
in a Chern-Simons model for the vector multiplet.

It is not difficult to generalize the construction given.
Suppose we have a system of $n+1$ Abelian vector multiplets, 
and let $W_a(z)$ be the corresponding field strengths, where $a=0,1,\dots,n$.
Instead of the single composite object (\ref{Gij}), we now have $(n+1)(n+2)/2$
such superfields defined as (compare with \cite{K,KT-M}) 
\bea
G^{ij}_{ab}:=\ri\,\cD^{\hga (i}W_a\cD_\hga^{j)} W_b+{\ri\over 2} W_{(a} \cD^{ij} W_{b)}
-2S^{ij}W_a W_b~, \qquad \cD^{(i}_\hga G^{jk)}_{ab}=0~.
\label{Gijab}
\eea
Assume also that  the field strength $W_0$ is everywhere non-vanishing, 
$\langle W_0 \rangle \neq 0$. Now, we can generalize eq. (\ref{Chern-Simons})
as follows:
\be
M^{ab}  G^{ij}_{ab} (z) =0~, \qquad M^{ab} =M^{ba}~,
\ee
where $M^{ab} $ is a constant nonsingular real  matrix 
normalized as  $M^{00}=1$.


\sect{Projective supermultiplets}

So far, we have provided the superspace realization for the main 
kinematic constructions of the superconformal tensor calculus \cite{Ohashi,Bergshoeff}. 
In the remainder of this paper, we  present new results that have not appeared in  
the component approaches of \cite{Ohashi,Bergshoeff}. 
We first introduce a large family of new off-shell (matter) 
supermultiplets coupled to the Weyl multiplet of 5D $\cN=1$ conformal
supergravity. They can be viewed to be a curved superspace generalization of  
the  known off-shell multiplets in 4D $\cN=2$ flat projective superspace 
\cite{KLR,LR,G-RLRvUW} or, more precisely, of the rigid 5D $\cN=1$ 
superconformal multiplets  \cite{K}.
Their off-shell structure is almost identical to that 
of the supermultiplets coupled to the minimal supergravity multiplet, which 
we have proposed in \cite{KT-Msugra5D}. 
Therefore, below we will closely follow \cite{KT-Msugra5D} and  specifically emphasize 
those features that are characteristic of conformal supergravity. 

In addition to the superspace coordinates $z^{\hat{M}}=(x^{\hm},\q^{\hat{\mu}}_i)$,
it is useful to introduce 
isotwistor  variables $u^{+}_i \in  
{\mathbb C}^2 \setminus  \{0\}$  defined to be inert with respect to 
the local group SU(2)  \cite{KT-Msugra5D}. 
The operators $\cD^+_{\hat \a}:=u^+_i\,\cD^i_{\hat \a} $  obey the following algebra:
\bea
\{  \cD^+_{\hat \a} , \cD^+_{\hat \b} \}
=-4{\rm i}\, \Big(X_{\hal \hbe}+N_{\hal \hbe}\Big)\,J^{++}
+4{\rm i} \, S^{++}M_{\hal \hbe}~,
\label{analyt}
\eea
where 
$J^{++}:=u^+_i u^+_j \,J^{ij}$ and 
$S^{++}:=u^+_i u^+_j \,S^{ij}$. 
Eq. (\ref{analyt}) follows from (\ref{covDev2spinor-}). 
It is tempting to consider constrained  superfields $Q(z,u^+)$ 
obeying the  constraint $\cD^+_{\hat \a} Q =0$.  For the latter  
to be consistent,  $Q(z,u^+)$  must be scalar with respect to the Lorentz group,
$ M_{\hal \hbe} Q=0$, and also possess special properties with respect to the
group SU(2), that is,  $J^{++}Q=0$. 
Let us  define such supermultiplets.

A projective supermultiplet of weight $n$,
$Q^{(n)}(z,u^+)$, is a scalar superfield that 
lives on  $\cM^{5|8}$, 
is holomorphic with respect to 
the isotwistor variables $u^{+}_i $ on an open domain of 
${\mathbb C}^2 \setminus  \{0\}$, 
and is characterized by the following conditions:\\
(i) it obeys the covariant analyticity constraint\footnote{In the case of rigid $\cN=2$ 
supersymmetry in four dimensions, 
similar constraints
were first introduced by Rosly 
\cite{Rosly},   and later by the harmonic \cite{GIKOS} 
and projective \cite{KLR,LR} superspace practitioners.}  
\be
\cD^+_{\hat \a} Q^{(n)}  =0~;
\label{ana}
\ee  
(ii) it is  a homogeneous function of $u^+$ 
of degree $n$, that is,  
\be
Q^{(n)}(z,c\,u^+)\,=\,c^n\,Q^{(n)}(z,u^+)~, \qquad c\in \mathbb{C}^*~;
\label{weight}
\ee
(iii) infinitesimal gauge transformations (\ref{tau}) act on $Q^{(n)}$ 
as follows:
\bea
\d_K Q^{(n)} 
&=& \Big( K^{\hat{C}} \cD_{\hat{C}} + K^{ij} J_{ij} \Big)Q^{(n)} ~,  
\non \\ 
K^{ij} J_{ij}  Q^{(n)}&=& -\frac{1}{(u^+u^-)} \Big(K^{++} D^{--} 
-n \, K^{+-}\Big) Q^{(n)} ~, \qquad 
K^{\pm \pm } =K^{ij}\, u^{\pm}_i u^{\pm}_j ~,
\label{harmult1}   
\eea 
where
\bea
D^{--}=u^{-i}\frac{\partial}{\partial u^{+ i}} ~,\qquad
D^{++}=u^{+ i}\frac{\partial}{\partial u^{- i}} ~.
\label{5}
\eea
The transformation law (\ref{harmult1}) involves an additional isotwistor,  $u^-_i$, 
which is subject 
to the only condition $(u^+u^-) = u^{+i}u^-_i \neq 0$, and is otherwise completely arbitrary.
By construction, $Q^{(n)}$ is independent of $u^-$, 
i.e. $\pa  Q^{(n)} / \pa u^{-i} =0$,
and hence $D^{++}Q^{(n)}=0$.
One can see that $\d Q^{(n)} $ 
is also independent of the isotwistor $u^-$, $\pa (\d Q^{(n)})/\pa u^{-i} =0$,
due to (\ref{weight}). 
It follows  from (\ref{harmult1})
\bea
J^{++} \,Q^{(n)}=0~, \qquad J^{++} \propto D^{++}~,
\label{J++}
\eea
and hence the covariant analyticity constraint (\ref{ana}) is indeed consistent.

In conformal supergravity, the important issue is how  
the projective multiplets may consistently vary under the super-Weyl transformations. 
Suppose we are 
given a weight-$n$ projective superfield $Q^{(n)}$ 
that transforms homogeneously, $\d_\s Q^{(n)}\propto  \s Q^{(n)}$. 
Then, its transformation law turns out to be  uniquely fixed 
by the constraint  (\ref{ana}).
\bea
\d_\s Q^{(n)}=3n \,\s Q^{(n)}~.
\label{super-Weyl-Qn}
\eea
The super-Weyl weight, $3n$,  
matches the superconformal weight of a rigid superconformal projective 
multiplet \cite{K} to which $Q^{(n)}$ reduces in the flat superspace limit. 
Without the assumption of homogeneity, 
it is easy to construct examples of projective multiplets which do not respect 
(\ref{super-Weyl-Qn}).  For instance, 
It follows from (\ref{S-Bianchi}) that $S^{++}$ is a projective superfield of weight two, 
\be
\cD^+_{\hat \a} S^{++}  =0~.
\ee
In accordance with (\ref{s-Weyl-Sij}), its super-Weyl transformation is
\be
\d_\s S^{++}=2\s S^{++}
+{\ri\over 2}\,(\cD^+)^2\s~, \qquad 
(\cD^+)^2:=\cD^{+\hal}\cD^+_{\hal}~.
\label{s-Weyl-S++}
\ee

Given a projective multiplet $Q^{(n)}$,
its complex conjugate 
is not covariantly analytic.
However, similarly to the flat four-dimensional case  
\cite{Rosly,GIKOS,KLR} , 
one can introduce a generalized,  analyticity-preserving 
conjugation, $Q^{(n)} \to \widetilde{Q}^{(n)}$, defined as
\be
\widetilde{Q}^{(n)} (u^+)\equiv \bar{Q}^{(n)}\big(
\overline{u^+} \to 
\widetilde{u}^+\big)~, 
\qquad \widetilde{u}^+ = {\rm i}\, \s_2\, u^+~, 
\ee
with $\bar{Q}^{(n)}(\overline{u^+}) $ the complex conjugate of $Q^{(n)}$.
Its fundamental property is
\bea
\widetilde{ {\cD^+_{\hat \a} Q^{(n)}} }=(-1)^{\e(Q^{(n)})}\, \cD^{+\hat \a}
 \widetilde{Q}{}^{(n)}~. 
\eea
One can see that
$\widetilde{\widetilde{Q}}{}^{(n)}=(-1)^nQ^{(n)}$,
and therefore real supermultiplets can be consistently defined when 
$n$ is even.
In what follows, $\widetilde{Q}^{(n)}$ will be called the smile-conjugate of 
${Q}^{(n)}$.

Important examples of projective supermultiplets are given in \cite{KT-Msugra5D}, 
and we refer the reader to that paper for more details.

Let $W$ be the field strength of an Abelian vector multiplet.
We can then introduce 
\be
G^{++} := G^{ij}u^+_iu^+_j
=\ri\,\cD^{+ \hal}W\cD_\hal^{+} W+{\ri\over 2} W (\cD^{+})^2 W-2S^{++}W^2~, 
\label{G++}
\ee
with $G^{ij}$ defined in (\ref{Gij}).
It follows from (\ref{G-anal}) that $G^{++}$ is a projective superfield of weight two, 
\be
\cD^+_{\hat \a} G^{++}  =0~.
\ee
In accordance with (\ref{super-Weyl-G}), the super-Weyl transformation of 
$G^{++}$  conforms with (\ref{super-Weyl-Qn}), 
\bea
\d_\s G^{++}=6 \s G^{++}~.
\label{super-Weyl-G++}
\eea

Consider a given supergravity background. The superconformal 
group of this space is defined to be generated by those combined
infinitesimal transformations  (\ref{tau}),  (\ref{sW1}) and  (\ref{sW2})
which do not change the covariant derivatives,
\be
\d_K \cD_{\hat A} + \d_\s \cD_{\hat A} =0~.
\ee
This definition is analogous to that often used in 4D $\cN=1$ supergravity
\cite{BK}.
In the case of 5D $\cN=1$ flat superspace, it is equivalent to the definition 
of the superconformal Killing vectors \cite{K}. In this case, the transformation laws 
of the projective multiplets reduce to those describing 
the rigid superconformal projective multiplets \cite{K}.

In defining the projective supermultiplets, 
we have used the reduced formulation for the Weyl multiplet. 
It is not difficult to see that this definition remains valid 
if the superspace geometry is realized in terms of the 
 formulation for the Weyl multiplet presented in section 2.
 Indeed, 
from (\ref{covDev2spinor-}) one deduces the anti-commutation relation:
\bea
\{  \cD^+_{\hat \a} , \cD^+_{\hat \b} \}
=-4{\rm i}\, \Big(X_{\hal \hbe}+N_{\hal \hbe}\Big)\,J^{++}
+4{\rm i} \, S^{++}M_{\hal \hbe}
-{\ri\over 2}\ve^{\ha\hb\hc\hd\he}(\S_{\ha\hb})_{\hal\hbe}C_{\hc}{}^{++}M_{\hd\he}
~,
\label{analyt-C}
\eea
where $C_{\ha}{}^{++}: = C_{\ha}{}^{ij} u^+_iu^+_j$.
It implies that the constraint (\ref{ana}) is consistent under the same 
conditions on $Q^{(n)}(z,u^+)$ which we have specified above.


\sect{Action principle}

Let $\cL^{++}$ be a real projective multiplet of weight two,  
and $W$  the  field strength of  a vector multiplet such that $\langle W \rangle \neq 0$.
We assume that $\cL^{++}$ possesses the 
following super-Weyl transformation:
\bea
\d_\s \cL^{++}=6\s \cL^{++}~
\label{super-Weyl-L++}
\eea
which complies with (\ref{super-Weyl-Qn}).
Associated with $\cL^{++}$  is the following functional 
\bea
S(\cL^{++})&=&
\frac{2}{3\pi} \oint (u^+ \rd u^{+})
\int \rd^5 x \,{\rm d}^8\q\,E\, \frac{\cL^{++}\,W^4}{(G^{++})^2}~, 
\qquad E^{-1}={\rm Ber}\, (E_{\hat A}{}^{\hat M})~.
\label{InvarAc}
\eea
This functional is  invariant under arbitrary re-scalings
$u_i^+(t)  \to c(t) \,u^+_i(t) $, 
$\forall c(t) \in {\mathbb C}\setminus  \{0\}$, 
where $t$ denotes the evolution parameter 
along the integration contour.
We are going to demonstrate  that $S(\cL^{++})$ 
does not change  under the  supergravity gauge transformations
and is super-Weyl invariant.
Therefore, eq. (\ref{InvarAc}) constitutes a locally supersymmetric 
and super-Weyl invariant action principle.

To prove the invariance of $S(\cL^{++})$
under infinitesimal  supergravity gauge transformations
(\ref{tau}) and (\ref{harmult1}),
we first point out that  
\be
Q^{(-2)}:=\frac{ \cL^{++} }{ (G^{++})^2}
\ee
 is a projective multiplet of weight $-2$,
because both $\cL^{++}$ and $G^{++}$ are projective multiplet of weight $+2$.
Since $W$ is SU(2)-scalar and $u$-independent,  from  eq. (\ref{harmult1})
we can deduce (see also \cite{KT-Msugra5D2})
\bea
K^{ij} J_{ij} \,\Big( Q^{(-2)}W^4\Big) = -\frac{1}{(u^+u^-)} D^{--}\Big( K^{++} Q^{(-2)}W^4\Big)~.
\eea
Next, since $K^{++} Q^{(-2)}$ has weight zero, it is easy to see 
\bea
(u^+ \rd u^{+} )\, K^{ij} J_{ij}  \,\Big(Q^{(-2)}W^4\Big) = -{\rm d}t \,
\frac{{ \rm d}  }{{\rm d}t} \, \Big(Q^{(-2)}W^4\Big)~, 
\eea
with  $t$ the evolution parameter along the integration contour in 
(\ref{InvarAc}). Since the integration contour is closed, the SU(2)-part of 
the transformation (\ref{harmult1}) does not contribute to the variation of 
the action (\ref{InvarAc}). To complete the proof of local supersymmetry invariance,
 it remains to take into the account 
the fact that $\cL^{++} /(G^{++})^2$  and $W$ are  Lorentz scalars.

To prove the invariance of $S(\cL^{++})$ under the infinitesimal super-Weyl 
transformations, we first note the transformation law of $E$:
\bea
\d_\s E=-2\s E~.
\eea
Now, it  only remains to take into account the transformation laws 
(\ref{super-Weyl-W}), (\ref{super-Weyl-G++}) and (\ref{super-Weyl-L++}).

Let us introduce the following fourth-order operator\footnote{This operator 
was considered for the first time in \cite{KT-M}
in the case of 5D $\cN=1$ anti-de Sitter supersymmetry. }
(see also \cite{KT-Msugra5D2}):
\bea
\D^{(+4)} = ({\cD}^+)^4 -\frac{5}{12} {\rm i}\, S^{++}\,({\cD}^+)^2
+3 (S^{++})^2~,
\label{anapro}
\eea
where 
\bea
(\cD^+)^4:=-{1\over 96}\ve^{\hal\hbe\hga\hde}
\cD^+_{\hal} \cD^+_{\hbe}\cD^+_{\hga}\cD^+_{\hde}~ .
\eea
Its crucial property is that the superfield $Q^{(n)} $ defined by
\bea
Q^{(n)} (z,u^+) := \D^{(+4)} U^{(n-4)} (z,u^+) ~, 
\eea
is a weight-$n$ projective multiplet, 
\bea
\cD^+_{\hat \a}Q^{(n)}=0~,
\eea
for any  {\it unconstrained} scalar superfield  $U^{(n-4)} (z,u^+) $  that
lives on  $\cM^{5|8}$, 
is holomorphic with respect to 
the isotwistor variables $u^{+}_i $ on an open domain of 
${\mathbb C}^2 \setminus  \{0\}$, 
and is characterized by the following conditions:\\
(i) it is  a homogeneous function of $u^+$ 
of degree $n-4$, that is,  
\be
U^{(n-4)}(z,c\,u^+)\,=\,c^{n-4}\,U^{(n-4)}(z,u^+)~, \qquad c\in \mathbb{C}^*~;
\ee
(iii) infinitesimal gauge transformations (\ref{tau}) act on $U^{(n-4)}$ 
as follows:
\bea
\d_K U^{(n-4)} 
&=& \Big( K^{\hat{C}} \cD_{\hat{C}} + K^{ij} J_{ij}  \Big)U^{(n-4)} ~,  
\non \\ 
K^{ij} J_{ij}  \,U^{(n-4)}&=& -\frac{1}{(u^+u^-)} \Big(K^{++} D^{--} 
-(n-4) \, K^{+-}\Big) U^{(n-4)} ~. \eea 
We will call $U^{(n-4)}(z,u^+)$ a {\it projective prepotential} 
for $Q^{(n)} $. 
In can be checked that 
$U^{(n-4)}$
should possess the super-Weyl transformation
\be
\d_\s U^{(n-4)} = (3n-4) \s U^{(n-4)}
\label{s-Weyl-U}
\ee
in order  for $Q^{(n)} =  \D^{(+4)} U^{(n-4)} $ to transform as in (\ref{super-Weyl-Qn}).

The important result is 
\bea
\D^{(+4)}W^4
={3\over 4}(G^{++})^2
~.
\label{idW}
\eea
This relation  can be proved  by using the identity 
\bea
\cD^+_\hal\cD^+_\hbe\cD^+_\hga W=-2\ri\ve_{\hbe\hga}S^{++}\cD^+_\hal W~
\label{3devW}
\eea
which follows from the Bianchi identity (\ref{W-BI-2}) with the aid of 
(\ref{analyt}).

Let $U^{(-2)}$ be a projective prepotential for the Lagrangian $\cL^{++}$ 
in (\ref{InvarAc})
\bea
\cL^{++}=\D^{(+4)}U^{(-2)}~.
\eea
Using the rule for integration by parts
\be
\int \rd^5 x \,{\rm d}^8\q\,E\, \cD_{\hat A} \, \F^{\hat A} =0~, 
\ee
for an arbitrary superfield $ \F^{\hat A} =
(\F^{\hat{a}}, \F^{\hat{\a}}_i) $, 
we obtain
\bea
\frac{2}{3\pi} 
\oint (u^+\rd u^{+})
\int \rd^5 x \,{\rm d}^8\q\,E\, \frac{\cL^{++}W^4}{(G^{++})^2}=
\frac{1}{2\pi} \oint (u^+\rd u^{+})
\int \rd^5 x \,{\rm d}^8\q\,E\, U^{(-2)}~,
\label{InvarAc2}
\eea
where we have used (\ref{idW}).
This crucial relation tells us that the supersymmetric action, 
eq. (\ref{InvarAc}),  is independent of the concrete choice 
of a vector multiplet with $\langle W \rangle \neq 0$, 
provided $\cL^{++}$ is independent of this vector multiplet.

It is worth pointing out that the super-Weyl invariance of the right-hand side in (\ref{InvarAc2}) 
also follows from (\ref{s-Weyl-U}).

Since the action  (\ref{InvarAc}) is super-Weyl invariant, 
one can choose the super-Weyl gauge (\ref{W=1}). 
Then, due to the explicit form of $G^{++}$, eq. (\ref{G++}), 
the action reduces to the functional 
\bea
S(\cL^{++})&=&
\frac{1}{6\pi} \oint (u^+ \rd u^{+})
\int \rd^5 x \,{\rm d}^8\q\,E\, \frac{\cL^{++}}{(S^{++})^2}
\eea
proposed in \cite{KT-Msugra5D2}.
As demonstrated  in \cite{KT-Msugra5D2},
this functional is a natural extension of the 
action principle in flat projective superspace \cite{KLR,S}.

It is useful to give several examples of supergravity-matters systems.
Let  ${\mathbb V}(z,u^+)$ denote the tropical prepotential for the central
charge vector multiplet appearing in the action  (\ref{InvarAc}) (see \cite{KT-Msugra5D}
for more detail).  It is a real weight-zero projective multiplet 
possessing the gauge invariance 
\be
\d {\mathbb V} =  \l +\tilde{\l}~, 
\ee
with $\l$ a weight-zero arctic multiplet  (see \cite{KT-Msugra5D}
for the definition of arctic multiplets).  A hypermultiplet can be described 
by an arctic weight-one multiplet 
$\U^{+ } (z,u^+) $ and its smile-conjugate 
$ \widetilde{\U}^{+}$.  Consider a  gauge invariant Lagrangian of the form 
(with the gauge transformation of $\U^+$ being $\d\U^+=-{\xi}\l\U^+$)
\be
\cL^{++} = \frac{1}{k^2} {\mathbb V} \,G^{++} 
-  \widetilde{\U}^{+} {\rm e}^{\x {\mathbb V} }\, \U^{+ }~,
\ee
with $\k$ Newton's constant, and $\x$ a cosmological constant. 
It describes Poincar\'e supergravity if $\x=0$, and 
pure gauge supergravity with $\x\neq0$.

A system of  arctic weight-one multiplets 
$\U^{+ } (z,u^+) $ and their smile-conjugates
$ \widetilde{\U}^{+}$ can be   described by the Lagrangian
\bea
\cL^{++} = {\rm i} \, K(\U^+, \widetilde{\U}^+)~,
\label{conformal-sm}
\eea
with $K(\F^I, {\bar \F}^{\bar J}) $ a real analytic function
of $n$ complex variables $\F^I$, where $I=1,\dots, n$.
${}$For $\cL^{++}$ to be a weight-two real projective superfield, 
it is sufficient to  require 
\bea
 \F^I \frac{\pa}{\pa \F^I} K(\F, \bar \F) =  K( \F,   \bar \F)~.
 \label{Kkahler2}
 \eea
This is a curved superspace generalization of the general model 
for  superconformal polar multiplets
\cite{K,KT-M,K2}.

As a generalization of the model given in \cite{KT-Msugra5D}, 
a system of interacting arctic weight-zero multiplets 
${\bf \U}  $ and their smile-conjugates
$ \widetilde{ \bf{\U}}$ can be described by the Lagrangian 
\bea
\cL^{++} = G^{++}\,
{\bf K}({\bf \U}, \widetilde{\bf \U})~,
\eea
with ${\bf K}(\F^I, {\bar \F}^{\bar J}) $ a real function 
which is not required to obey any 
homogeneity condition. 
The action is invariant under K\"ahler transformations of the form
\be
{\bf K}({\bf \U}, \widetilde{\bf \U})~\to ~{\bf K}({\bf \U}, \widetilde{\bf \U})
+{\bf \L}({\bf \U}) +{\bar {\bf \L}} (\widetilde{\bf \U} )~,
\ee
with ${\bf \L}(\F^I)$ a holomorphic function.
\\


\noindent
{\bf Acknowledgements:}\\
This work is supported  in part
by the Australian Research Council and by a UWA research grant.


\small{

}


\begin{thebibliography}{66}

\bibitem{KT-Msugra5D}
S.~M.~Kuzenko and G.~Tartaglino-Mazzucchelli,
  ``Five-dimensional superfield supergravity,''
    Phys.\ Lett.\ B {\bf 661}, 42 (2008),
 [arXiv:0710.3440].

\bibitem{KT-Msugra5D2}
  S.~M.~Kuzenko and G.~Tartaglino-Mazzucchelli,
  ``5D supergravity and projective superspace,''
  JHEP {\bf 0802}, 004 (2008) [arXiv:0712.3102].

\bibitem{Howe5Dsugra}
  P.~S.~Howe,
 ``Off-shell N=2 and N=4 supergravity in five-dimensions,''
in: M. J. Duff and C. J. Isham (Eds.),
 {\it Quantum Structure of Space and Time}, 
 Cambridge Univ. Press, 
1982, p. 239.

\bibitem{HL}
  P.~S.~Howe and U.~Lindstr\"om,
  ``The supercurrent in five dimensions,''
  Phys.\ Lett.\ B {\bf 103}, 422 (1981).

\bibitem{GST}
  M.~G\"unaydin, G.~Sierra and P.~K.~Townsend,
  ``The geometry of N=2 Maxwell-Einstein supergravity and Jordan algebras,''
  Nucl.\ Phys.\  B {\bf 242}, 244 (1984); 
``Gauging the D = 5 Maxwell-Einstein supergravity theories: More on Jordan
 algebras,''
  Nucl.\ Phys.\  B {\bf 253}, 573 (1985).

\bibitem{GZ}
  M.~G\"unaydin and M.~Zagermann,
  ``The gauging of five-dimensional, N = 2 Maxwell-Einstein supergravity
  theories coupled to tensor multiplets,''
  Nucl.\ Phys.\  B {\bf 572}, 131 (2000)
  [hep-th/9912027].

\bibitem{CD}
  A.~Ceresole and G.~Dall'Agata,
  ``General matter coupled N = 2, D = 5 gauged supergravity,''
  Nucl.\ Phys.\  B {\bf 585}, 143 (2000)
  [hep-th/0004111].

\bibitem{Ohashi}
  T.~Kugo and K.~Ohashi,
  ``Off-shell d = 5 supergravity coupled to matter-Yang-Mills system,''
  Prog.\ Theor.\ Phys.\  {\bf 105}, 323 (2001)
   {[hep-ph/0010288]};
  T.~Fujita and K.~Ohashi,
  ``Superconformal tensor calculus in five dimensions,''
  Prog.\ Theor.\ Phys.\  {\bf 106}, 221 (2001)
 {[hep-th/0104130]}.

\bibitem{Bergshoeff}
 E.~Bergshoeff, S.~Cucu, M.~Derix, T.~de Wit, R.~Halbersma and A.~Van Proeyen,
  ``Weyl multiplets of N = 2 conformal supergravity in five dimensions,''
  JHEP {\bf 0106}, 051 (2001)
  [hep-th/0104113];
  E.~Bergshoeff, S.~Cucu, T.~de Wit, J.~Gheerardyn, R.~Halbersma, 
  S.~Vandoren and A.~Van Proeyen,
  ``Superconformal N = 2, D = 5 matter with and without actions,''
  JHEP {\bf 0210}, 045 (2002)
  {[hep-th/0205230]};
E.~Bergshoeff, S.~Cucu, T.~de Wit, J.~Gheerardyn, S.~Vandoren and A.~Van Proeyen,
  ``N = 2 supergravity in five dimensions revisited,''
  Class.\ Quant.\ Grav.\  {\bf 21}, 3015 (2004)
 {[hep-th/0403045]}. 
 
 \bibitem{Zucker}
  M.~Zucker,
``Minimal off-shell supergravity in five dimensions,''
  Nucl.\ Phys.\ B {\bf 570}, 267 (2000)
{[hep-th/9907082]};
``Gauged N = 2 off-shell supergravity in five dimensions,''
  JHEP {\bf 0008}, 016 (2000)
{[hep-th/9909144]};
``Off-shell supergravity in five-dimensions and supersymmetric 
brane world scenarios,''
  Fortsch.\ Phys.\  {\bf 51}, 899 (2003).

\bibitem{GIKOS}
A.~S.~Galperin, E.~A.~Ivanov, S.~N.~Kalitsyn, V.~Ogievetsky, E.~Sokatchev, 
``Unconstrained N=2 matter, Yang-Mills and supergravity theories in harmonic
superspace,''
Class.\ Quant.\ Grav.\  {\bf 1}, 469 (1984).
 
\bibitem{GIOS}
A.~S.~Galperin, E.~A.~Ivanov, V.~I.~Ogievetsky and E.~S.~Sokatchev,
{\it Harmonic Superspace}, Cambridge University Press,  2001.

\bibitem{KT}
 S.~M.~Kuzenko and S.~Theisen,
  ``Correlation functions of conserved currents in N = 2 superconformal
  theory,''
  Class.\ Quant.\ Grav.\  {\bf 17}, 665 (2000)
  [hep-th/9907107].

\bibitem{BW}
J.~Bagger and E.~Witten,
``Matter couplings in N=2 supergravity,''
Nucl.\ Phys.\  B {\bf 222}, 1 (1983).

\bibitem{Swann} A. Swann, ``HyperK\"ahler and quaternionic K\"ahler geometry,''
Math. Ann. {\bf 289}, 421 (1991).

\bibitem{Galicki}
  K.~Galicki,
  ``Geometry of the scalar couplings in N=2 supergravity models,''
  Class.\ Quant.\ Grav.\  {\bf 9}, 27 (1992).

\bibitem{deWRV}
  B.~de Wit, M.~Ro\v{c}ek and S.~Vandoren,
  ``Hypermultiplets, hyperk\"ahler cones and quaternion-K\"ahler geometry,''
  JHEP {\bf 0102}, 039 (2001)
  [hep-th/0101161].
  
\bibitem{KLR}
A. Karlhede, U. Lindstr\"om and M. Ro\v cek,
``Self-interacting tensor multiplets in N=2 superspace,''
Phys.\ Lett.\ B {\bf 147}, 297 (1984).

\bibitem{LR}
  U.~Lindstr\"om and M.~Ro\v{c}ek,
 ``New hyperk\"ahler  metrics  and new supermultiplets,''
  Commun.\ Math.\ Phys.\  {\bf 115}, 21 (1988);
  ``N=2 super Yang-Mills theory in projective superspace,''
  Commun.\ Math.\ Phys.\  
  {\bf 128}, 191 (1990).

\bi{G-RLRvUW}
F.~Gonzalez-Rey, U.~Lindstr\"om
M.~Ro\v{c}ek, R.~von Unge and S.~Wiles, 
 ``Feynman rules in N = 2 projective superspace. 
I: Massless  hypermultiplets,''
  Nucl.\ Phys.\ B {\bf 516}, 426 (1998) {[hep-th/9710250]}.
 
 \bibitem{K}
 S.~M.~Kuzenko,
 ``On compactified harmonic/projective superspace, 5D superconformal
 theories, and all that,''
  Nucl.\ Phys.\  B {\bf 745}, 176 (2006)
  [hep-th/0601177]. 
  
  \bibitem{K2}
S.~M.~Kuzenko, ``On superconformal projective hypermultiplets,''
JHEP {\bf 0712}, 010 (2007) [arXiv:0710.1479].
  
\bibitem{KLRT}
S. M. Kuzenko, U. Lindstr\"om, M. Ro\v{c}ek
and G. Tartaglino-Mazzucchelli, 
``4D  $\cN=2$ Supergravity and Projective Superspace,''
work in progress.  
  
\bibitem{NR}
  H.~Nishino and S.~Rajpoot,
  ``Alternative N = 2 supergravity in five dimensions with singularities,''
  Phys.\ Lett.\  B {\bf 502}, 246 (2001)
  [arXiv:hep-th/0011066].
  
\bibitem{Zupnik}
  B.~Zupnik,
  ``Harmonic superpotentials and symmetries in gauge theories with eight
  supercharges,''
  Nucl.\ Phys.\  B {\bf 554}, 365 (1999)   [Erratum-ibid.\  B {\bf 644}, 405 (2002)]
  [hep-th/9902038].

\bibitem{KL}
  S.~M.~Kuzenko and W.~D.~Linch, III,
  ``On five-dimensional superspaces,''
  JHEP {\bf 0602}, 038 (2006)
  [hep-th/0507176].

\bibitem{Howe}
  P.~S.~Howe,
  ``Supergravity in superspace,''
  Nucl.\ Phys.\  B {\bf 199}, 309 
(1982).

\bibitem{Muller} M. M\"uller, {\it Consistent Classical Supergravity Theories},
(LectureNotes in Physics, Vol. 336),
Springer, Berlin, 1989. 
  
\bi{KT-M}
S.~M. Kuzenko and G. Tartaglino-Mazzucchelli,
``Five-dimensional N=1 AdS superspace:
Geometry,  off-shell multiplets and dynamics,''
Nucl. Phys. B {\bf 785}, 34 (2007),
 [arXiv:0704.1185].
  
\bi{Rosly}
 A.~A.~Rosly,
``Super Yang-Mills  constraints 
as integrability conditions,'' in {\it Proceedings of the International 
Seminar on Group Theoretical 
Methods in Physics},'' (Zvenigorod, USSR, 1982),
M. A. Markov  (Ed.), 
Nauka, Moscow, 1983, Vol. 1, p. 263 (in Russian);
English translation: in {\it Group Theoretical 
Methods in Physics},'' M. A. Markov, V. I. Man'ko 
and A. E. Shabad  (Eds.), Harwood Academic Publishers, 
London, Vol. 3, 1987, p. 587.

\bibitem{BK}
I.~L. Buchbinder and S.~M. Kuzenko, {\it Ideas and Methods of Supersymmetry and
Supergravity, Or a Walk Through Superspace}, IOP, Bristol, 1998.

\bibitem{S}
W. Siegel,
``Chiral actions for N=2 supersymmetric tensor multiplets"
Phys.\ Lett.\ B {\bf 153} (1985) 51. 


\end{thebibliography}
\end{document}